# The Age of Sensorial Zero Trust: Why We Can No Longer Trust Our Senses

**Fabio Correa Xavier**

fabio@fabioxavier.com.br

www.fabioxavier.com.br

www.linkedin.com/in/fabiocorreaxavier

## Abstract

In a world where deepfakes and cloned voices are emerging as sophisticated attack vectors, organizations require a new security mindset: Sensorial Zero Trust [9]. This article presents a scientific analysis of the need to systematically doubt information perceived through the senses, establishing rigorous verification protocols to mitigate the risks of fraud based on generative artificial intelligence. Key concepts, such as Out-of-Band verification, Vision-Language Models (VLMs) as forensic collaborators, cryptographic provenance, and human training, are integrated into a framework that extends Zero Trust principles to human sensory information. The approach is grounded in empirical findings and academic research, emphasizing that in an era of AI-generated realities, even our eyes and ears can no longer be implicitly trusted without verification. Leaders are called to foster a culture of methodological skepticism to protect organizational integrity in this new threat landscape.

## 1 Introduction

The increasing sophistication of generative artificial intelligence (GenAI) has fundamentally transformed the cybersecurity landscape, ushering in an era where the very reality perceived by our senses becomes a critical vulnerability. Previously, the implicit trust in the authenticity of a familiar voice or a recognizable face was a cornerstone of human and organizational interactions. Today, technologies such as deepfakes and AI-cloned voices – often created via Generative Adversarial Networks (GANs) – enable the convincing fabrication of reality with unprecedented ease and accessibility. These forgeries include techniques like faceswap, reenactment, and synthetic generation, which directly challenge the human capacity to distinguish the real from the fabricated [31]. The quality of such manipulations is now so high that they can deceive even the most discerning eyes and ears. The proliferation of false information and high-impact financial fraud provides robust empirical evidence that trust, once an asset, has become a critical vector of vulnerability. For instance, an executive's voice can be imitated to steal hundreds of thousands of dollars via a simple phone call. Research indicates an explosive 10-fold (1000%) global increase in detected deepfake incidents between 2022 and 2023, with Brazil experiencing an 830% surge and North America a staggering 1740% increase [34]. Globally, GenAI-enabled fraud is projected to cost Brazil roughly R$4.5 billion in 2025, and Gartner analysts predict that by 2025 half of all phishing attacks will leverage deepfakes or synthetic voices [34].

Moreover, recent studies reveal a significant cognitive vulnerability: 70% of individuals are not confident in their ability to distinguish a real voice from a cloned one [36]. In Brazil, a Kaspersky survey found that 66% of the population does not even know what a deepfake is, making them easy targets for social engineering [39]. Beyond these threats, GenAI multi-agent systems (MAS) introduce novel attack vectors like "tool squatting" – the deceptive registration or misrepresentation of software tools – which can lead to data exfiltration, resource abuse, system compromise, and a profound erosion of trust within the MAS ecosystem [4]. Notably, insiders can perpetrate such attacks if tool registries are uncontrolled, and AI models themselves can be co-opted for model-based identity theft by replicating behavioral patterns of legitimate agents. In response to this evolving threat landscape, we propose the paradigm of Sensorial Zero Trust.

This concept extends the already established Zero Trust Architecture (ZTA), which operates under the principle of "never trust, always verify." Historically, ZTA emerged around 2010 when John Kindervag of Forrester Research formalized a shift away from the old "castle-and-moat" model (where everything inside the network perimeter was implicitly trusted) towards a model of zero implicit trust. Kindervag's Zero Trust framework posits that no entity – internal or external – should be automatically trusted without verification, effectively "de-perimeterizing" network security [2].

The U.S. National Institute of Standards and Technology (NIST), in Special Publication 800-207, formally defines Zero Trust as "a cybersecurity paradigm focused on resource protection and the premise that trust is never granted implicitly but must be continuously evaluated" [3].

In practice, a Zero Trust Architecture is an enterprise security plan that continuously authenticates and authorizes based on all available data points, enforcing least-privilege access and monitoring for anomalies [1]. Sensorial Zero Trust adapts these principles to the domain of human perception and interaction with AI-mediated content, requiring systematic and methodical doubt regarding information received through the senses – especially when digitally mediated [1]. It necessitates rigorous verification protocols based on several pillars, which we introduce below.

By integrating technological and human-centered defenses, this approach aims to ensure that no sensory input is trusted by default [1]. In the sections that follow, we detail the vulnerabilities inherent in AI-mediated sensory perception and outline a framework for implementing Sensorial Zero Trust. Ultimately, organizational leaders are challenged to re-evaluate their processes and culture for an environment in which, as science demonstrates, sensory information can no longer be trusted without rigorous verification.

## 2. The Theoretical Foundation of Zero Trust

### 2.1 Conceptualization and Principles

Zero Trust is not a single technology but a strategic philosophy born from the dissolution of the traditional network perimeter [8]. Under the old paradigm, once a user or device was inside the network, it was often granted broad, implicit trust [1]. This perimeter-based "trust by default" approach has proved dangerously inadequate in modern environments of cloud services and remote work [1]. In 2010, John Kindervag popularized the term "Zero Trust," advocating to "never trust, always verify" for every user, device, and transaction [2]. In essence, Zero Trust operates on the assumption that systems will inevitably be compromised, and therefore no entity—whether inside or outside the network—should be automatically trusted [30]. Every access request must be authenticated, authorized, and continuously validated before being allowed. As one recent study succinctly put it, assuming any actor is safe by virtue of network location is a "risky assumption", and a proper Zero Trust approach "insists that no device, user, or system be trusted by default". This philosophy was formally codified by NIST in SP 800-207.

According to NIST, "Zero Trust assumes no implicit trust is granted to assets or user accounts based solely on their physical or network location", and trust must be earned through continual verification of identity, context, and policy adherence. A Zero Trust Architecture therefore embeds several core principles of cybersecurity: Verify Explicitly – always authenticate and authorize based on all available data points (user identity, device health, location, etc.), Use Least-Privileged Access – grant the minimum access needed to perform an action, and Assume Breach – design defenses such that a breach is contained through micro-segmentation and continuous monitoring.

In practice, this means that every request for a resource is treated as untrusted until proven otherwise, no matter where it originates from or what it claims to be [1]. Academic research corroborates the effectiveness of this model. For example, a multivocal literature review by Buck et al. emphasizes that Zero Trust requires thorough authentication and does not automatically trust any asset or user based on network location [7]. Similarly, Nasiruzzaman et al. note that implicit trust based on being "inside" the network is a fundamental weakness of traditional models, and that modern security must continuously verify every access request given the ever-blurring network boundaries [1]. the Zero Trust paradigm establishes that trust is not a status, but a continuous process – always conditional, always subject to re-evaluation. These foundational ideas set the stage for extending Zero Trust beyond networks and devices, into the realm of human sensory information.

**2.2 Fundamental Pillars of Zero Trust**

Building on NIST guidelines and scholarly analyses, the core pillars of Zero Trust can be outlined as follows:

(i) **Verify Explicitly**: Always authenticate and authorize each request, using all available signals. Identity must be confirmed, device integrity checked and context (such as time, geolocation and behavioral signals) evaluated every time access is requested. No access is granted solely because a request comes from within a corporate LAN or a known IP range – location is not identity. For example, even an internal user must pass multi-factor authentication and device health checks to access an internal resource. This philosophy of "never trust, always verify" underpins Zero Trust and mandates continuous verification of all access requests [1][7].

(ii) **Least-Privileged Access**: Limit access rights for users and devices to the bare minimum needed to perform their duties, following "just-in-time" and "just-enough-access" principles. By curtailing broad or standing privileges, Zero Trust mitigates the damage that can occur if an account is compromised. A user compromise should only grant the attacker access to a narrow slice of resources, not the entire network. Implementing least privilege reduces the attack surface and prevents lateral movement within compromised networks [1].

(iii) **Assume Breach**: Operate as if an intrusion has already occurred. Design network segments and resource access with the expectation that any single component could be compromised. This mindset leads to strong micro-segmentation (isolating resources into secure zones) and robust encryption of all communications by default. Continuous monitoring and anomaly detection are employed to catch malicious activity that evades preventive controls. Treating every environment change as a potential breach vector ensures readiness through rigorous risk assessment and incident response planning [3][5]

These pillars form the foundation of Zero Trust security. They represent a shift from one-time verification at login to continuous verification and validation at every step. If implemented properly, any request for data or system access is treated with healthy skepticism. As a result, even if attackers penetrate one layer of defense—by stealing credentials or breaching a device—additional checks and limited privileges stop them from freely moving laterally across the IT environment. This model has proven effective: organizations adopting Zero Trust have reduced incident impacts and limited insider threats, according to several industry surveys and case studies. In the context of Sensorial Zero Trust, these same pillars must apply to information coming through our eyes and ears. Just as Zero Trust in networks means verifying every device and user, Zero Trust in sensory data means verifying every video, voice or image before trusting it. The next sections describe the technological deceptions that force us to doubt our senses and how we can systematically extend Zero Trust principles to mitigate these emerging threats.

# 3. The Technologies of Deception: Scientific Fundamentals

### 3.1 Deepfakes – Generative Adversarial Networks (GANs)

The primary threat justifying a Sensorial Zero Trust stance is the rise of deepfakes—hyper-realistic fake images or videos, often of people, generated by AI [35]. At the core of most deepfake generation are Generative Adversarial Networks (GANs), a class of machine-learning frameworks where two neural networks are pitted against each other. One network, the Generator, creates synthetic images (or videos or audio), while the other, the Discriminator, evaluates whether the output is real or fake. Through millions of iterative rounds, the Generator improves to the point that its forgeries become exceptionally high-quality and convincing [35]. This GAN architecture enables the creation of "convincing representations of public figures, corporate executives, or even family members" with relatively minimal input data. For example, a GAN can study a few photographs of a person and produce entirely new images or videos of that person saying and doing things they never did. The Discriminator network's feedback guides the Generator to focus on and correct any telltale flaws. Over time, the forgeries become so realistic that humans struggle to tell them apart from authentic content [32]. Research surveys have catalogued common deepfake methods—from face swaps (mapping one person's face onto another's body), to facial reenactment (altering facial movements to change expressions or speech), to full synthetic generation (creating a completely fictional face) [35]. Each method presents unique detection challenges. Deepfakes can be weaponized for disinformation, propaganda, or impersonation in social engineering attacks [40], making them a critical threat to trust in visual evidence.

### 3.2 Voice Cloning – The State of the Art

Analogous advancements have occurred in AI-based voice cloning. Modern voice cloning techniques use deep learning to analyze a person's unique vocal features—tone, accent, cadence, inflection, even breathing patterns—and then reproduce that voice saying arbitrary phrases with only a short sample of someone's speech [32]. An encoder network first captures the essential characteristics of the voice into a representation, and a decoder then applies those characteristics to new text (text-to-speech) or to another voice recording (voice style transfer). The result is the target voice speaking words it never actually said.

Recent research reveals alarming progress in this field. A 2024 study by Barrington et al. (published in Scientific Reports) demonstrated that, with only a few seconds of audio, today's commercial tools can produce digital voice replicas that fool people a large fraction of the time [32]. In that study, human listeners correctly identified a voice as AI-generated only about 60 % of the time, and they mistakenly perceived the identity of an AI-generated voice to be the same as the real person approximately 80 % of the time [32]. These results underline that even trained individuals are barely better than chance at detecting short cloned voice clips. Another experiment found that just three seconds of sample audio can enable an 85 % voice match to the original speaker [36].

The implications for security are profound. Voice is often used as an authentication factor (think of a bank authorizing transfers by a voice call) or as a trust signal ("I recognized my boss's voice on the phone, so I followed the instructions"). With AI voice cloning, that trust signal can be counterfeit. Indeed, documented incidents show attackers impersonating CEOs' voices to trick subordinates into authorizing large fund transfers—a UAE-based energy firm lost €220 000 in one such voice scam in 2019 [33]. As these tools become even more accessible (some are offered as inexpensive cloud services or apps), organizations can no longer assume that a familiar voice on a call or a face in a video feed is legitimate without additional verification.

The technical landscape of deepfakes and voice clones has advanced to the point that our innate human senses are increasingly unreliable detectors of forgery. Vision and hearing, once the bedrock of verifying reality, have become vulnerable to manipulation by AI. This necessitates a shift in mindset: just as Zero Trust networking treats any packet or login as potentially malicious until proven safe, Sensorial Zero Trust treats any visual or auditory content as potentially fake until verified. Before turning to solutions, we first quantify the risk with real-world cases and data.

## 4. Risk Analysis: Paradigmatic Cases and Quantitative Impact

### 4.1 Documented Paradigm Cases
Several high-profile fraud cases illustrate the paradigm shift we are experiencing:

- **The Hong Kong Deepfake Heist (2024):** In this incident, a bank manager in Hong Kong was deceived into authorizing approximately US $ 25.6 million in transfers. The attackers set up a sophisticated multi-person video conference where all the other participants—purportedly business executives—were AI-generated deepfakes. The visual deepfakes, combined with social engineering tactics (creating a false sense of urgency and legitimacy), were so convincing that the manager did not suspect fraud until after the money was gone

[40]. This case is a chilling milestone: it represents perhaps the first known use of multiple deepfakes in a live business meeting to execute fraud.
- **The UAE Voice-Cloning Scam (2020):** Criminals used an AI voice clone to impersonate the voice of a company director and then called that company's bank manager, urgently requesting a transfer. Believing he was speaking with his client, the manager approved a US $ 35 million transfer to the scammers' account. The clone even replicated the director's slight accent and manner of speaking [33]. This demonstrates how a few seconds of sampled voice, coupled with urgent pretexts, can defeat the human safeguards in financial transactions.
- **The European Energy Firm Fraud (2019):** One of the earliest known voice deepfake scams occurred when fraudsters mimicked a German CEO's voice. They called the CEO's subordinate in a UK-based subsidiary and convinced him that an emergency payment was needed. The subordinate, hearing what he thought was his boss's voice—complete with the familiar German accent—complied and transferred € 220 000 to the attackers [33]. This case, reported in 2019, foreshadowed the larger wave of AI-aided frauds that would follow.

These cases underscore a common theme: traditional trust cues were exploited and undermined by AI forgeries. The victims relied on voices and faces as authenticators – a natural thing to do, since historically only a person's real face or voice could appear live on a call. That assumption is no longer valid. With deepfakes, seeing is not necessarily believing; with voice clones, hearing is not believing.

**4.2 Epidemiological Data of the Threat**

Beyond individual anecdotes, data from studies and industry reports show the alarming scale and trajectory of AI-enabled deception:

- **Explosive Growth**: Sumsub, a digital identity verification company, documented a 1000 % global increase in detected deepfake incidents from 2022 to 2023 [34]. In Brazil, the growth was about 830 % year-over-year, while North America saw an even steeper rise of 1740 %. This surge reflects both increased usage of deepfakes by bad actors and improved detection capabilities revealing a latent epidemic.
- **Economic Impact Projections**: Gartner analysts estimate that by 2025, 50 % of all phishing attacks will incorporate deepfake or synthetic voice elements. In monetary terms, one industry report forecasted that AI-based fraud (including deepfakes) would cost organizations worldwide over $50 billion annually in the near term, with Brazil alone possibly incurring R$ 4.5 billion in losses by 2025.
- **Cognitive Vulnerability**: A McAfee survey covering several countries found that 70 % of people doubted their ability to tell a cloned voice from a real one [36]. Meanwhile, in Brazil, a Kaspersky poll revealed two-thirds of the population (66 %) have no awareness of what deepfakes are [39]. Such lack of awareness and confidence means many individuals are ill-prepared to question what they see and hear, which is exactly what scammers exploit. Social engineering tactics become far more effective when victims trust the "evidence" of their senses, even if that evidence is falsified.

Taken together, these cases and statistics make it clear that the threat of AI-driven deception is not theoretical or limited – it is already here and proliferating. Trust, in its traditional form, has become a liability. To address this, organizations must operationalize distrust of sensory information through concrete policies and technologies. In the next section, we outline how to implement Sensorial Zero Trust as a methodological approach to counter these threats.

# 5. Implementing Sensorial Zero Trust: A Methodological Approach

To mitigate the risks outlined above, Sensorial Zero Trust demands multi-layered controls that align with Zero Trust principles. We describe four key pillars of implementation: Out-of-Band Verification, Extended Authentication, Continuous Monitoring, and Automated Deepfake Detection – all supported by strong awareness training.

**5.1 Out-of-Band (OOB) Verification**

Out-of-Band verification is a foundational pillar of Sensorial Zero Trust. It mandates using a separate, independent communication channel to confirm any request received through a primary channel [38]. In practice, this means whenever a sensitive action is initiated via one medium, a secondary medium is used to verify authenticity. For example, if a wire transfer request comes by email or video call, an OOB policy would require confirmation via a different channel — such as a phone call to a pre-registered number or an in-person verification. This approach adds a critical layer of security, as an attacker would need to compromise two independent channels simultaneously to succeed [38].

OOB verification is particularly effective against deepfake and impersonation scams because it capitalizes on the difficulty of an adversary mimicking a target consistently across disparate media. An imposter might convincingly fake an executive's voice on a phone call, but they would have much more trouble also intercepting that executive's personal cell phone or reproducing a consistent identity over a video call with live interaction. By breaking the path of a single-mode attack, OOB checks can catch attempts that would otherwise go through [38]. Indeed, cybersecurity experts strongly advocate OOB authentication (OOBA) as an essential defense for high-risk transactions, noting that it provides a second barrier if one factor (like an email) is compromised. Many banks have long used this (e.g. confirming large transfers via a text message code); Sensorial Zero Trust extends it to any scenario where decisions are made based on potentially spoofable sensory information.

**5.2 Extended Multi-Factor Authentication (MFA)**

Multi-factor authentication – requiring more than one piece of evidence (factor) to verify identity – is already a mainstream security control for logins. Under Sensorial Zero Trust, MFA should be extended beyond just user logins to include critical actions and communications [3]. For instance, before executing an unusual financial transaction requested over a video conference, an employee might require the requester to authenticate via a second factor (like approving a push notification on their phone or quoting a one-time code).

Essentially, treat certain sensitive verbal or written requests as you would an account login – subject to MFA [3]. While MFA isn't infallible (attackers have developed techniques like MFA fatigue or real-time phishing of OTP codes to try to bypass it), those tend to rely on tricking the user, reinforcing the need for layered defenses and user training [7].

The key point is that MFA significantly raises the effort bar for attackers. In a Sensorial Zero Trust model, MFA isn't just for logging into systems, but for confirming identities in person or in real-time communications. For example, an organization might issue physical or digital identity tokens to executives; if an employee receives an urgent instruction from a supposed executive via voice or video, they could challenge the instruction with a second-factor check (e.g., "Please confirm this request by responding to the authentication app notification I just sent you"). Only if the true executive approves on their device (which the fraudster wouldn't possess) will the request be treated as legitimate [3].

**5.3 Continuous Authentication and Behavioral Biometrics**

Traditional authentication is a point-in-time event – e.g., enter password (and maybe a code) once, and you're in. Continuous authentication aims to verify the user on an ongoing basis, in the background, by analyzing behavioral and physiological signals. Sensorial Zero Trust can leverage continuous authentication to detect anomalies that might indicate an ongoing session has been hijacked or that the person behind an action isn't who they claim [3]. For example, systems can monitor behavioral biometrics such as typing patterns, mouse movements, touchscreen gestures, gait (from phone motion sensors) or voice timbre over a call.

These behaviors are hard to perfectly mimic. If a deepfake video is being streamed, it might mimic a person's face and voice, but it cannot replicate how the real person types or swipes on their device. If the purported user's behavior deviates significantly from their historical profile, the system can raise an alert or re-prompt for verification [7]. Prior research has found that these biometric patterns can be distinctive enough to identify individuals continuously. For instance, keystroke dynamics can achieve high accuracy in distinguishing users, and combining multiple behavioral signals can improve confidence [7].

Continuous identity checks align with Zero Trust's "never trust" ethos by not assuming that a session should remain trusted after initial login [3]. In the context of human communication, one could imagine future conferencing software that continuously analyzes voice patterns to ensure they match the enrolled user's voiceprint, potentially catching when an AI impostor takes over the audio stream.

Similarly, if an employee's known typing cadence suddenly changes drastically during a chat conversation where they're receiving unusual requests (perhaps because a hacker or AI is generating the responses), an automated system could flag this. While continuous authentication for general scenarios is still an evolving field, incorporating these emerging technologies can significantly bolster a Sensorial Zero Trust framework by providing ongoing validation that "person X is still person X."

### 5.4 Automated Deepfake Detection Technologies

Human senses alone will not reliably catch deepfakes, but technology can assist. An array of automated detection tools has emerged, aiming to spot the subtle artifacts and inconsistencies that often escape the human eye or ear [35]:

- **Liveness detection**: Common in the biometrics industry, liveness detection algorithms check if a presented face or voice is from a live person and not a recording or deepfake. For example, liveness detection in video calls might ask the user to turn their head or blink at random intervals—something a static deepfake video would struggle with. Advanced versions analyze micro-movements of skin or eyeballs and the 3D characteristics of faces to ensure what's on camera is a live human being [7]. In audio, liveness tests might involve asking the speaker to repeat random phrases or perform challenges that a pre-recorded deepfake couldn't anticipate. These measures are increasingly used in remote onboarding processes (e.g., when opening a bank account via a mobile app, you may be asked to take a selfie video with liveness checks to prevent someone from using a photo of you). Integrating liveness checks into high-stakes communications can add confidence that there's a real person on the other side [7].
- **Deepfake content analysis**: Researchers have developed specialized detection algorithms that analyze media files for signs of tampering. For instance, deepfake videos might exhibit visual artifacts—irregularities in lighting, skin texture or edge consistency—especially in freeze frames. There are known giveaways like unnatural eye-blinking patterns or slight distortions in facial accessories (glasses, earrings) across frames. On the audio side, generated voices may have spectral artifacts or lack the dynamic range of a human voice. Cutting-edge detectors, often based on neural networks themselves, can be trained on large datasets of fakes vs. reals to recognize these subtle cues [35]. Some tools focus on specific inconsistencies: for example, checking if head movements and facial expressions align with expected physiological patterns, or analyzing whether eye reflections match the environment (deepfake algorithms sometimes fail to render realistic eye reflections). While no detector is foolproof—adversaries adapt and some deepfakes can bypass many tests—using multiple detection techniques in tandem greatly increases the chance of exposure. It's analogous to antivirus: a deepfake detector might not catch everything, but it will catch many known attack patterns and raise the cost for an attacker to produce an undetectable fake [35].

In implementing Sensorial Zero Trust, organizations should deploy these detection mechanisms at key checkpoints. For example, a company could require that any recorded video or audio submitted as part of a process (a video interview, a voice instruction for a transaction, etc.) be scanned by deepfake detection software, with any positives flagged for additional human verification. Similarly, live video-conferencing platforms used for critical meetings could have built-in optional deepfake scanning modes. It's important to note that detection should not be relied on alone—it complements process changes like OOB verification and MFA described earlier. But technology can shoulder the initial burden of filtering out obvious fakes and issuing warnings, which is invaluable given the speed and scale at which attacks might come.

### 5.5 Developing Human Detection Competencies

While technology is essential, the human element remains a crucial line of defense. Employees and the general user base must be educated and trained to maintain a healthy skepticism and to recognize the hallmarks of deepfake or impersonation attempts [8, 10]. Just as classic security awareness training teaches people how to spot phishing emails (e.g., poor grammar, urgent pleas, mismatched URLs), modern training needs to cover deepfake indicators and social engineering red flags. Key components of a human-centric Sensorial Zero Trust training program include:

- **Psychological Awareness**: Attackers using deepfakes often rely on psychological tactics—creating a sense of urgency, secrecy or authority pressure—to get targets to act quickly without verifying. Training should reinforce that urgent requests—especially those involving financial transactions or sensitive data—must be verified, not taken at face value, even if they appear to come from a high-ranking person. Role-playing scenarios can be effective: for example, simulate a call where a "CEO" orders an emergency fund transfer, then walk through the verification steps the employee should take (and highlight the consequences of not doing so). Emphasize that no legitimate leader will fault an employee for taking the time to verify a sensitive request—in fact, leadership should explicitly encourage this behavior [8].
- **Technical Tells of Deepfakes**: Equip staff to recognize common deepfake artifacts. In video, watch for irregular blinking or a staring, unvarying gaze; overly smooth skin lacking subtle microexpressions; lip-sync errors where audio and mouth movements fall out of step; and inconsistent lighting or shadows—such as earrings that blur or merge into the skin between frames. In audio, listen for mechanical intonation, unnaturally timed pauses or pacing, and incongruous background noise (for example, completely silent ambience or a repeating noise loop). Although these cues become more subtle as deepfake technology advances, a trained eye or ear can still detect fleeting glitches that even automated detectors flag [35]. Encourage employees also to ask spontaneous, unscripted questions during live interactions—deepfake systems often stumble when forced to depart from pre-recorded responses.
- **Encouraging Reporting and Second Opinions**: Foster a culture where employees feel comfortable pausing and escalating if something "feels off." If an employee suspects a voice or video might be fake (or simply unusual), they should have a clear, non-punitive path to report it or get a second opinion. This could be an emergency security hotline or a procedure like "call back the person via a known official number." The important part is that employees don't dismiss their own doubts. Many social engineering scams succeed because the target had a fleeting doubt but overrode it to be polite or obedient. Sensorial Zero Trust culture should validate and reward doubt. Leadership can set the tone by sharing examples: e.g., a CEO might tell staff, "If you ever get a message that seems like it's from me asking for something odd, please verify through another channel. I will never be upset at that – on the contrary, I expect it" [3].

Technology and human training must reinforce each other. Automated tools might detect what humans miss, and humans might notice context or content oddities that tools haven't been tuned to catch. Together, they create a more robust shield. By instituting both advanced technical controls and comprehensive awareness programs, organizations move closer to a true Sensorial Zero Trust posture – one in which every sensory input is scrutinized, verified, and validated through multiple means before trust (even provisional) is granted.

## 6. Leadership in the Age of Artificial Reality

Implementing a Sensorial Zero Trust strategy is not just a technical or procedural challenge; it is fundamentally a leadership imperative. Executive leaders and managers play a pivotal role in shifting organizational culture from implicit trust to constructive skepticism.

### 6.1 The Mandate of the CISO

The Chief Information Security Officer (CISO) and their team will naturally be at the forefront of this initiative. The role of the CISO has evolved dramatically in recent years – from primarily a technical administrator to a strategic business leader [3]. In the context of Sensorial Zero Trust, the CISO must articulate the business risks posed by deepfakes and AI-driven fraud in terms that the Board and C-suite understand, quantifying potential financial losses, brand damage and legal liabilities from a successful deepfake-enabled attack [7]. By translating these risks into business impact, the CISO can build the case for investments in necessary training, processes and technologies.

One critical leadership task is to update incident response plans and policies to account for AI-mediated threats. For example, a CISO should ensure that the company's fraud response playbooks include procedures for suspected deepfake incidents (e.g., steps to validate communications, involve forensic experts, etc.) [5]. They also need to institutionalize some of the controls discussed, like OOB verification, by issuing clear policies – for instance: "No wire transfer above $X should be executed without voice confirmation via a known number," or "Any video meeting requesting disclosure of confidential information must be recorded and subject to verification before action" [6]. These policies must be backed by top leadership so that employees feel empowered (and obliged) to follow them even if it means slowing things down.

The CISO should champion simulation exercises ("deepfake drills") similar to how companies run phishing tests [3]. By periodically testing the organization with benign deepfake scenarios, the security team can gauge preparedness and reinforce learning. Ultimately, the CISO's mandate in Sensorial Zero Trust is to weave a web of trust that is multi-dimensional—covering people, processes and technology—and to constantly adapt it as threats evolve.

### 6.2 Leading by Example

A robust security culture starts at the very top [3]. Leaders must visibly model the behaviors that Sensorial Zero Trust requires. If the CEO or other senior executives bypass verification steps for the sake of convenience, it sends a message that security protocols are optional or can be ignored when inconvenient, undermining the cultural change needed [7]. Conversely, when employees see leaders meticulously following the same rules—for example, a CFO who insists an employee call them back on a known number to confirm a request, even though it was really the CFO on the first call—it reinforces the legitimacy and importance of the controls [7].

Leaders should also openly discuss the threats. Company-wide communications might include a note from the CEO like: "In today's world of AI, I have instructed our team and I expect you to verify unusual requests even if they appear to come from me. If you get a voicemail from 'me' asking for something strange, double-check it. I will do the same with any of you" [3]. This type of messaging does two things: it raises awareness and it grants permission to doubt higher authorities when appropriate. It essentially says doubting is not disrespecting; doubting can be protecting.

On an operational level, organizations might consider a "secure communications" policy for executives. This could involve agreed-upon channels for sensitive communications (for instance, a secure corporate messenger that has strong encryption and in-app biometric identity verification) so that if an executive suddenly sends instructions over an unofficial channel, it will automatically raise skepticism. Leaders should adhere to using these official channels themselves. An example might be: if the policy is "we only discuss wire transfer approvals via our secure app or in-person," then the CEO should never try to shortcut it by sending a casual email or text for an approval—doing so would normalize the behavior attackers are hoping for [3].

Leaders must be the champions and exemplars of Sensorial Zero Trust, because culture change flows downward [7].

## 7. Integration with Broader Security Frameworks

Sensorial Zero Trust is not a standalone concept in competition with other frameworks, but rather an extension and enhancement of a comprehensive cybersecurity strategy. It aligns with and builds upon existing models such as the NIST Cybersecurity Framework and Zero Trust Architecture guidelines. NIST SP 800-207 (Zero Trust Architecture) emphasizes a "constant cycle of obtaining access, scanning and assessing threats, adapting, and continually reevaluating trust in ongoing communications". Sensorial Zero Trust takes this constant cycle and applies it to the domain of human perception – essentially adding new "threat signals" (like potential deepfake indicators) into the loop of assessment and response. Organizations should integrate Sensorial Zero Trust measures incrementally, targeting use-cases where the risk is highest. For example, integrate deepfake detection and verification steps first in processes like vendor payment authorizations, CEO communications, and customer identity verifications, where a successful impersonation would be catastrophic. This use-case approach mirrors recommendations in existing frameworks to tackle highest risks first and evolve iteratively. Sensorial Zero Trust also connects with identity management and authentication frameworks. Decades of work in identity and access management (IAM) are now being augmented by concepts like Decentralized Identifiers (DIDs) and Verifiable Credentials (VCs) to support trust in distributed environments [6, 22]. These can be important in verifying sources of information. For instance, imagine a future where every legitimate video a company produces is cryptographically signed by the media software using the company's private key – any unsigned video purporting to be from that company can then be treated as suspect. This is akin to code-signing certificates, but for multimedia content authenticity. Such ideas are being explored in standards bodies (e.g., W3C's Verifiable Credentials Data Model). Sensorial Zero Trust will benefit from and contribute to these developments. An example is the emerging concept of "authentication of provenance" for media: ensuring that an email, document, or video can be traced to a verified source. By incorporating provenance verification tools (some of which leverage blockchain or distributed ledgers for integrity), organizations can more confidently automate the trust/distrust decisions on incoming media. Another area of integration is with Incident Response (IR) and Cyber Resilience plans. Sensorial Zero Trust should be reflected in IR plans by preparing playbooks for deepfake incidents (e.g., "What do we do if someone releases a fake video of our CEO making false statements to tank our stock or brand? What if employees receive malicious deepfake calls that appear to be from IT asking for passwords?"). These scenarios need to be anticipated and drilled. This ties into business continuity and crisis communication plans as well. It's a cross-functional effort: Legal, PR, and HR might all need to be involved in responding to certain deepfake scenarios (like reputational attacks or harassment via fake videos of employees). By integrating such playbooks, an organization is better prepared to limit damage when prevention fails [20]. Sensorial Zero Trust doesn't replace existing frameworks but fills a critical gap in them. Traditional Zero Trust tells us to verify devices, identities, and transactions; Sensorial Zero Trust adds "verify your eyes and ears when it matters." Together, they form a more complete defense. Adopting Sensorial Zero Trust measures will naturally overlap with improvements in identity security, threat intelligence (since staying ahead of deepfake trends will be part of threat intel), and user training regimes. The result is a more holistic and robust security posture that addresses both digital and cognitive vulnerabilities.

## 8. Final Considerations

Sensorial Zero Trust is not a standalone concept in competition with other frameworks, but rather an extension and enhancement of a comprehensive cybersecurity strategy [2It aligns with and builds upon existing models such as the NIST Cybersecurity Framework [9] and Zero Trust Architecture guidelines. NIST SP 800-207 (Zero Trust Architecture) emphasizes a 'constant cycle of obtaining access, scanning and assessing threats, adapting, and continually reevaluating trust in ongoing communications' [3]. Sensorial Zero Trust takes this constant cycle and applies it to the domain of human perception—essentially adding new "threat signals" (like potential deepfake indicators) into the loop of assessment and response.

Organizations should integrate Sensorial Zero Trust measures incrementally, targeting use-cases where the risk is highest. For example, integrate deepfake detection and verification steps first in processes like vendor payment authorizations, CEO communications and customer identity verifications, where successful impersonation would be catastrophic. This use-case approach mirrors recommendations in existing frameworks to tackle highest risks first and evolve iteratively [9, 10].

Sensorial Zero Trust also connects with identity management and authentication frameworks. Decades of work in identity and access management (IAM) are now being augmented by concepts like Decentralized Identifiers (DIDs) [22] and Verifiable Credentials (VCs) [6] to support trust in distributed environments. These can be important in verifying sources of information. For instance, imagine a future where every legitimate video a company produces is cryptographically signed by the media software using the company's private key—any unsigned video purporting to be from that company can then be treated as suspect. This is akin to code-signing certificates, but for multimedia content authenticity. Such ideas are being explored in standards bodies (e.g., W3C's Verifiable Credentials Data Model) [6]. Sensorial Zero Trust will benefit from and contribute to these developments. An example is the emerging concept of "authentication of provenance" for media: ensuring that an email, document or video can be traced to a verified source [23]. By incorporating provenance verification tools (some of which leverage blockchain or distributed ledgers for integrity) [18], organizations can more confidently automate the trust/distrust decisions on incoming media.

Another area of integration is with Incident Response (IR) and Cyber Resilience plans. Sensorial Zero Trust should be reflected in IR plans by preparing playbooks for deepfake incidents (e.g., "What do we do if someone releases a fake video of our CEO making false statements to tank our stock or brand? What if employees receive malicious deepfake calls that appear to be from IT asking for passwords?") [13]. These scenarios need to be anticipated and drilled. This ties into business continuity and crisis communication plans as well. It's a cross-functional effort: Legal, PR and HR might all need to be involved in responding to certain deepfake scenarios (like reputational attacks or harassment via fake videos of employees).

By integrating such playbooks, an organization is better prepared to limit damage when prevention fails. Sensorial Zero Trust doesn't replace existing frameworks but fills a critical gap in them. Traditional Zero Trust tells us to verify devices, identities and transactions; Sensorial Zero Trust adds "verify your eyes and ears when it matters." Together, they form a more complete defense. Adopting Sensorial Zero Trust measures will naturally overlap with improvements in identity security, threat intelligence (since staying ahead of deepfake trends will be part of threat intel [24]) and user training regimes. The result is a more holistic and robust security posture that addresses both digital and cognitive vulnerabilities.

https://www.cnnbrasil.com.br/business/golpistas-usam-deepfake-de-diretor-financeiro-e-roubam-us-25-milhoes/